\documentclass[a4paper, 12pt]{article}
\usepackage{latexsym}
\usepackage{amssymb}

\begin{document}
\noindent{\large\bf A note on discrete monotonic dynamical
systems}

\vskip 1cm
{\small

\noindent{\bf Dongsheng Liu}\\

\noindent Department of Physics, Lancaster University, Lancaster,
LA1 4YB, UK\\
\noindent E-mail: d.liu@lancaster.ac.uk

\vskip .5cm \noindent {\bf Abstract.} We give a upper bound of
Lebesgue measure $V(S(f,h,\Omega))$ of the set $S(f,h,\Omega)$ of
points $q\in Q_h^d$ for which the triple $(h,q,\Omega)$ is
dynamically robust when $f$ is monotonic and satisfies certain
condition on some compact subset $\Omega \in \mathbb{R}^d$.}

\vskip .6cm \noindent{\bf 1. Introduction} \vskip .2cm

A discrete dynamical system on the state space $\mathbb{R}^d$ is generated by the iteration of a mapping $f: \mathbb{R}^d \to \mathbb{R}^d,$ that is $ x_{n+1}=f(x_n), n=0,1,2,\cdots.$\\

Let $Q_h^d$ denote the $h$-cube in $\mathbb{R}^d$ centered at origin, that is
$$ Q_h^d=\{x=(x_1,x_2,\cdots,x_d)\in \mathbb{R}^d : -\frac{h}{2}<x_i\leq \frac{h}{2}, i=1,2,\cdots,d\}$$
and for each $q \in Q_h^d$, let $L_{h,q}=\{q+hz : z \in \mathbb{Z}^d\}$ be the uniform $h$-lattice in $\mathbb{R}^d$ centered at $q$.\\

For $q \in Q_h^d$, we define the roundoff operator $[.]_{h,q}$ from $\mathbb{R}^d$ into $L_{h,q}$ by
$[x]_{h,q}=L_{h,q}\cap (x+Q_h^d)$ for $x \in \mathbb{R}^d$, or equivalently by
$$ [x]_{h,q}=([x_1-q_1]_h+q_1, \cdots, [x_d-q_d]_h+q_d)$$
where $x=(x_1,x_2,\cdots.x_d), q=(q_1,q_2,\cdots,q_d)$ and $[y]_h$ is scalar roundoff operator defined by
$$[y]_h=kh \qquad \textrm{ if } (k-\frac{1}{2})h\leq y < (k+\frac{1}{2})h.$$
Let $f$ be a dynamical system in $\mathbb{R}^d$. The map $f_{h,q}: L_{h,q} \to L_{h,q}$ defined by $$f_{h,q}(x)=[f(x)]_{h,q}, \qquad x\in L_{h,q}$$
is called $L_{h,q}$-discretization of $f$.

Now we give the definition of dynamical robustness \cite{pd}:\\
Given $h>0$, $q\in Q_h^d$, and a compact set  $\Omega \subset\mathbb{R}^d$, we say the triple $(h,q,\Omega)$ is dynamically robust if the discretization $f_{h,q}$ has a single equilibrium $x_{h,q}=f_{h,q}(x_{h,q})\in \Omega \cap L_{h,q}$ and
$$\lim_{n \to \infty}|f^n_{h,q}(x)-x_{h,q}|=0 \qquad  \forall x \in L_{h,q}\cap \Omega. $$

In \cite{pd} the following question was raised: given $f$ and a compact set $\Omega \in \mathbb{R}^d$, what is the Lebesgue measure $V(S(f,h,\Omega))$ of the set $S(f,h,\Omega)$ of points $q\in Q_h^d$ for which the triple $(h,q,\Omega)$ is dynamically robust?

They answered this question partially: when $\Omega$ is a parallel-polyhedron in $\mathbb{R}^d$ , $f$ is monotonic on $\Omega$ and satisfies some condition, they give a lower bound for $V(S(f,h,\Omega))$. In this paper we give a upper bound of $V(S(f,h,\Omega))$ for $f$ is monotonic and satisfies certain condition on some compact subset $\Omega \in \mathbb{R}^d$.

\vskip .4cm \noindent{\bf 2. Main Results} \vskip .2cm

We give the semi-ordering in $\mathbb{R}^d$: for $x,y \in \mathbb{R}^d$, we say $x\leq y$ if $x_i \leq y_i$ for $i=1,2,\cdots,d$ and $x < y $ if $x_i<y_i$ for $i=1,2,\cdots,d$. We shall say $f$ is monotonically increasing on a set $S\in \mathbb{R}^d$ if $f(x)\leq f(y)$ for all $x,y \in S$ with $x\leq y$.

In the following we restrict attention to monotonically increasing
functions, noting that the monotonic decreasing case is handled
similarly.

By the definition of dynamically robust, we have
\newtheorem{pro}{Proposition}
\begin{pro}
Let $f$ be monotonic on the compact set $\Omega\subset\mathbb{R}^d$ and suppose that $(h,q,\Omega)$ is a dynamically robust, $x_{h,q}$ is the single equilibrium. $\forall x \in L_{h,q}\cap \Omega$, if $x\ge x_{h,q}$, $f_{h,q}(x)=x_{h,q}$; if $x \leq x_{h,q}$, there exists $k\in \mathbb{N}$ such that $f_{h,q}^k(x)=x_{h,q}$.
\end{pro}
\emph{Proof.} Because $\Omega$ is a compact subset of $\mathbb{R}^d$, $L_{h,q}\cap \Omega$ is a finite set. For $x \in L_{h,q}\cap \Omega$, if $x \leq x_{h,q}$, let $f_{h,q}(x)=x^{(1)}$, then $x^{(1)}=f_{h,q}(x)\leq f_{h,q}(x_{h,q})=x_{h,q}$. If $x^{(1)}=x_{h,q}$ it is proved with $k=1$. Otherwise we consider $x^{(2)}:=f_{h,q}(x^{(1)})\leq f_{h,q}(x_{h,q})=x_{h,q}$. If $x^{(2)}=x_{h,q}$ it is proved with $k=2$. If $x^{(2)} \ne x_{h,q}$ we can continue this process. But $L_{h,q}\cap \Omega$ is finite set there exists a $k\in \mathbb{N}$ such that $f^k_{h,q}(x)=f_{h,q}(x^{(k-1)})=x_{h,q}$. If $x\ge x_{h,q}$, and $f_{h,q}(x)\ne x_{h,q}$, by the monotonicty, $f_{h,q}(x)\ge f_{h,q}(x_{h,q})=x_{h,q}$. so $|f^n_{h,q}(x)-x_{h,q}|\ge |f_{h,q}(x)-x_{h,q}|>0$ for any $n\in \mathbb{N}$, It is contradiction to the definition of dynamically robust of $(h,q,\Omega)$. So we have $f_{h,q}(x)=x_{h,q}$.

In fact, $\forall x \in L_{h,q}\cap \Omega$ there exists $k\in
\mathbb{N}$ such that $f_{h,q}^k(x)=x_{h,q}.$

\vskip.2cm
\noindent Now we can estimate $V(S(f,h,\Omega)).$
\newtheorem{th1}{Theorem}
\begin{th1}
$\Omega$ is a compact subset of $\mathbb{R}^d$ and satisfies: $\forall q \in Q_h^d$, there exist $u_{1,q}, u_{2,q} \in L_{h,q}\cap \Omega$ such that $\forall x \in L_{h,q}\cap \Omega$, $u_{1,q}\leq x\leq u_{2,q}$. Let $f$ be monotonic on the compact set $\Omega \subset \mathbb{R}^d$ and $f(\Omega)\subset \Omega'$ where $\Omega' \subset \Omega$ and satisfies $\forall x=(x_1,\cdots,x_d)\in \partial\Omega$, the boundary of $\Omega$, and $\forall x'=(x_1',\cdots,x_d')\in \Omega'$, $|x_i-x_i'|\ge \frac{h}{2}$, $i=1,2,\cdots,d.$ we have
$$V(S(f,h,\Omega))\leq \frac{L}{L-1}h^d-\frac{1}{L-1}V(\{x\in \Omega: x-\frac{h}{2}\leq f(x)< x+\frac{h}{2}\})$$
where $L=L_1\times L_2 \times \cdots \times L_d$ and $L_i$ is determined by following: let $l_i=|max\{x_i: x=(x_1,\cdots,x_i,\cdots,x_d)\in \Omega\}-min\{x_i: x=(x_1,\cdots,x_i,\cdots,x_d) \in \Omega\}|$ and $l_i=rh+p$, $0\leq p<h$ then $L_i=r+1$.
\end{th1}
\emph{Proof}. The method of this proof is following that in \cite{pd}. Let $F(h,q)=L_{h,q}\cap \{x: x-\frac{h}{2} \leq f(x)<x+\frac{h}{2}\} $ and $k(h,q)=\#\{ F(h,q)\}$. In order to carry on proof, we need following
\newtheorem{le1}{Lemma}
\begin{le1}
\cite{mg}.
$$\int_{Q_h^d}k(h,q)dq=V(\{x: x-\frac{h}{2}\leq f(x)<x+\frac{h}{2} \}).$$
\end{le1}
We also need the following special case of the Birkhoff-Tarski Theorem
\newtheorem{le2}[le1]{Lemma}
\begin{le2}
\cite{gb}. Let $g$ be a monotonic map of a finite set $\Gamma \in
\mathbb{R}^d$ into itself. If $g$ satisfies $g(x)\ge x$ or
$g(x)\leq x$ for $x \in \Gamma$, then the iterative sequence
$x_{n+1}=g(x_n)$ with $x_0=x$ converge to the fixed point
$g(x^*)=x^* \in \Gamma$.
\end{le2}
\emph{Remark}:\\
(1). We can get the fixed point by following: take any $x\in \Gamma$ with $g(x)\ge x$ or $g(x)\leq x$ and iterate $x_{n+1}=g(x_n)$ with $x_0=x$, because $\Gamma$ is finite, after a finite number of steps we can get a fixed point.\\
(2). If $f_{h,q}$ has only one fixed point $x^{*}$, then
$x^{*}=f_{h,q}^k(u_{1,q})$ and $x^{*}=f_{h,q}^l(u_{2,q})$ for some
$k,l \in \mathbb{N}$. Since $u_{1,q}\leq x\leq u_{2,q}$ it is easy
to see $f_{h,q}^n(x)=x^{*}$, $\forall x \in L_{h,q}\cap \Omega$
for large $n\in\mathbb{N}$.
\vskip.3cm
The condition on $f$
guarantees that $f_{h,q}$ is a mapping of $L_{h,q}\cap \Omega$
into itself. The elements of $F_{h,q}$ are precisely the fixed
points of $f_{h,q}$. So it is easy to see $k(h,q)\ge 1$ from the
lemma 2 because $f_{h,q}(u_{1,q})\ge u_{1,q}$. By the definition
of dynamically robust and remark (2), we have $q \in
S(f,h,\Omega)$ if and only if $k(h,q)=1$. So
$V(S(f,h,\Omega))=V(\{q: k(h,q)=1\})$. But
$$V(\{q: k(h,q)=1\})+V(\{q: k(h,q)>1\})=h^d,$$
and $k(h,q)$ at most equal to $L=L_1\times \cdots \times L_d$. By lemma 1, we have

$$V(\{x: x-\frac{h}{2}\leq f(x)<x+\frac{h}{2}\}) = \int_{Q_h^d} k(h,q)dq$$
\begin{eqnarray*}
&=& V(\{q: k(h,q)=1\})+\sum_{i=2}^L i\times V(\{q: k(h,q)=i\})\\
&\leq & V(\{q: k(h,q)=1\})+L\times V(\{q: k(h,q)>1\})\\
&=&V(\{q: k(h,q)=1\})+Lh^d-L\times V(\{q: k(h,q)=1\})\\
&=& Lh^d-(L-1)V(\{q: k(h,q)=1\})\\
&=& Lh^d -(L-1)V(S(f,h,\Omega)).
\end{eqnarray*}
So,
$$ V(S(f,h,\Omega))\leq \frac{L}{L-1}h^d-\frac{1}{L-1}V(\{x\in\Omega, x-\frac{h}{2}\leq f(x)< x+\frac{h}{2}\}).$$

Under the condition of Theorem 2, the result of Theorem 1 in
\cite{pd} still holds. Combining with  the Theorem 1 in \cite{pd},
we get
\newtheorem{co}{Corollary}
\begin{co}
Under the condition of Theorem 2 below we have\\
$Max \{ 0, 2h^d-V(\{x\in \Omega:x-\frac{h}{2}\leq f(x)< x+\frac{h}{2}\})\}\leq V(S(f,h,\Omega))\\
\leq  \frac{L}{L-1}h^d-\frac{1}{L-1}V(\{x \in \Omega, x-\frac{h}{2}\leq f(x)< x+\frac{h}{2}\}).$
\end{co}

\emph{Remark}: It is easy to see $h^d\leq V(\{x\in \Omega: x-\frac{h}{2}\leq f(x)<x+\frac{h}{2}\})\leq Lh^d.$

If $f$ is not monotonic, the situation is complex. Following we give a special example.

For $g$ is a map from $\Omega$ into itself, we say $x$ is a periodic point of $g$, if there exist $n\in \mathbb{N}$ such that $g^n(x)=x$. The least $n$  which satisfies $g^n(x)=x$ is called period of $g$ at $x$.

\vskip.2cm \noindent Now we give the example.
\newtheorem{th2}[th1]{Theorem}
\begin{th2}
Let $f$ be a map from a compact set $\Omega$ into $\Omega'$. where $\Omega' \subset \Omega$ and satisfies $\forall x=(x_1,\cdots,x_d)\in \partial\Omega$, the boundary of $\Omega$, and $\forall x'=(x_1',\cdots,x_d')\in \Omega'$, $|x_i-x_i'|\ge \frac{h}{2}$, $i=1,2,\cdots,d.$ If $\forall q\in Q_h^d$, $f_{h,q}$ has no periodic point with period more than $1$,  then
$$V(S(f,h,\Omega))\leq \frac{L}{L-1}h^d-\frac{1}{L-1}V(\{x\in \Omega: x-\frac{h}{2}\leq f(x)< x+\frac{h}{2}\})$$
where $L=L_1\times L_2 \times \cdots \times L_d$ and $L_i$ is determined by following: let $l_i=|max\{x_i: x=(x_1,\cdots,x_i,\cdots,x_d)\in \Omega\}-min\{x_i: x=(x_1,\cdots,x_i,\cdots,x_d) \in \Omega\}|$ and $l_i=rh+p$, $0\leq p<h$ then $L_i=r+1$.
\end{th2}
\emph{Proof}. We note $$V(\{q: k(h,q)=0\})+V(\{q:
k(h,q)=1\})+V(\{q: k(h,q)>1\})=h^d,$$ so $$V(\{q: k(h,q)>1\}) \leq
h^d-V(\{q: k(h,q)=1\}).$$ Now we only need to prove following.
\newtheorem{le3}[le1]{Lemma}
\begin{le3}
$$q \in S(f,h,\Omega) \qquad \textrm{if and only if}\quad  k(h,q)=1.$$
\end{le3}
\emph{Proof}. Let $q \in S(f,h,\Omega)$, but $k(h,q)\ne 1$. Then $k(h,q)=0$ or $k(h,q)>1$. That means dynamical system $f_{h,q}$ has no equilibrium or has at least two distinct equilibria, it is contradition to $q\in S(f,h,\Omega)$.

If $k(h,q)=1$, let $x_{h,q}$ is the unique fixed point of $f_{h,q}$ in $L_{h,q}\cap \Omega$. $\forall x_1 \in L_{h,q}\cap \Omega$, the condition of $f$ guarantee $f_{h,q}(x_1)\in L_{h,q}\cap \Omega$. Let $F_{h,q}(x_1):=x_2$, if $x_2 \ne x_1$, we consider $f_{h,q}(x_2):=x_3$. If $x_3 \ne x_2$ then $x_3 \ne x_1$ since $f_{h,q}$ has no periodic point with period more than 1. We continue this process and get $x_1,x_2, \cdots, \in L_{h,q}\cap \Omega$, which are pairwise distinct . But $L_{h,q}\cap \Omega$ is finite, so after finite number of steps, say $N$ steps, we have $f_{h,q}^N(x_1)=f_{h,q}(x_N)=x_N $. But $x_{h,q}$ is the unique fixed point, we get $x_N=x_{h,q}$ and $f_{h,q}^N(x_1)=x_{h,q}$. So $f^m_{h,q}(x_1)=x_{h,q}$ for any $m\ge N$. that is
$$\lim_{n \to \infty}f_{h,q}^n(x)=x_{h,q}, \qquad \forall x\in L_{h,q}\cap \Omega.$$
i.e., $q\in S(f,h,\Omega).$

The next step of the proof is the same as that in Theorem 2.

\end{document}